\documentstyle[12pt]{article}
\input epsf

\input epsf

\begin{document}

\renewcommand{\thesection}{\arabic{section}.} 
\renewcommand{\theequation}{\thesection \arabic{equation}}
\newcommand{\scs}{\setcounter{equation}{0} \setcounter{section}}
\def\req#1{(\ref{#1})}
\newcommand{\be}{\begin{equation}} \newcommand{\ee}{\end{equation}} 
\newcommand{\ba}{\begin{eqnarray}} \newcommand{\ea}{\end{eqnarray}} 
\newcommand{\la}{\label} \newcommand{\nb}{\normalsize\bf} 
\newcommand{\lb}{\large\bf} \newcommand{\vol}{\hbox{Vol}}
\newcommand{\bb} {\bibitem} \newcommand{\np} {{\it Nucl. Phys. }} 
\newcommand{\pl} {{\it Phys. Lett. }} 
\newcommand{\pr} {{\it Phys. Rev. }} \newcommand{\mpl} {{\it Mod. Phys. Lett. }}
\newcommand{\sg}{{\sqrt g}} \newcommand{\sqhat}{{\sqrt{\hat g}}}
\newcommand{\sqphi}{{\sqrt{\hat g}} e^\phi} 
\newcommand{\sqalpha}{{\sqrt{\hat g}}e^{\alpha\phi}}
\newcommand{\tp}{\cos px\ e^{(p-{\sqrt2})\phi}} \newcommand{\stwo}{{\sqrt2}}
\newcommand{\tr}{\hbox{tr}}

\begin{titlepage}
\renewcommand{\thefootnote}{\fnsymbol{footnote}}

\hfill CERN-TH/2000-151

\hfill hep-th/0005248

\vspace{.4truein}
\begin{center}
 {\LARGE Micrometer Gravitinos
\vskip3mm

 and the Cosmological Constant}
 \end{center}
\vspace{.7truein}

 \begin{center}

 Christof Schmidhuber\footnote{christof.schmidhuber@cern.ch}

 \vskip5mm

 {\it CERN, Theory Division, 1211 Gen\`eve 23}

 \end{center}

\vspace{.5truein}
\begin{abstract}

\noindent
We compute the 4--dimensional cosmological constant in string
compactifications in which the Standard Model fields live on
a non-supersymmetric brane inside a supersymmetric bulk.
The cosmological constant receives
contributions only from the vacuum energy of the bulk supergravity fields,
but not from the vacuum energy of the brane fields.
The latter is absorbed in a warp factor.
Supersymmetry breaking on the brane at the TeV scale
implies supersymmetry breaking in the bulk at
the micrometer scale. This creates a tiny cosmological constant that
agrees with experiment within a few orders of magnitude.
Our argument predicts superpartners of the graviton with mass  
of order $10^{-3}eV$. They
could be observed in short-distance tests of Einstein Gravity.

\end{abstract}
 \renewcommand{\thefootnote}{\arabic{footnote}}
 \setcounter{footnote}{0}
\end{titlepage}

\subsection*{1. Introduction}\scs{1}

The observed smallness of the cosmological constant $\lambda$ 
in Einstein's equations
poses a fine-tuning problem already in classical field theory
coupled to gravity. E.g., when the Higgs field rolls down its
potential, the energy density of the vacuum  and thus the effective
value of $\lambda$
changes by a large amount.
Other expected contributions to $\lambda$ that are suspiciously 
absent include those from condensates in QCD.

But perhaps the most mysterious aspect of the problem is that $\lambda$
does not seem to receive contributions from the quantum mechanical
ground state energies
$$\rho_k\ =\ {\hbar\omega\over2}\ \ \ ,\ \ \ \omega^2\ =\ k^2+m^2$$
of the oscillators with momentum $k$ of the 
massless and light fields of the Standard Model.
Summing
these contributions up to some large-momentum cutoff $\Lambda\gg m$,
one finds in the case of a single--component bosonic field \cite{wei}:
\ba
\lambda\ =\ 8\pi G\int_0^\Lambda {k^2dk\over2\pi^2}\rho_k\ \sim\ 
\ {\Lambda^4\over2\pi}\ l_P^2\ .\la{diane}\ea
Here, $G$ is the Newton constant and
$l_P={\sqrt{\hbar G}}\sim1.7\cdot10^{-35}m$ is the Planck length
(setting $c=1$). 
 Experimentally, it presently seems that \cite{bah}
\ba\lambda\ \sim\ (2\cdot 10^{-33}eV)^2\ \sim\ (10^{26}m)^{-2}\  
\la{flavia}\ea
(setting $\hbar=G=1$).
$10^{26}m$ is the order of magnitude of the curvature radius 
of the universe, which is roughly
the inverse Hubble constant. But even if one considered only the contributions
of the two helicity states
 of the massless photon to the cosmological constant (\ref{diane}), then in order to
explain such a small value of $\lambda$ one would need a momentum--cutoff
as small as
$$\Lambda\ \sim\ {1\over100}eV\ \sim\ \ 
 {1\over20 \mu m}\ .$$
So the minimum wavelength would have to
be as large as $20 \mu m$.
Already the wavelength of visible light is much smaller than 20 $\mu m$.
In this sense the observed smallness of the Hubble expansion parameter seems
inconsistent even with what we can see with our bare eyes.

Supersymmetry could explain a zero cosmological constant, because
the vacuum energies of the superpartners cancel each other.
Supersymmetry looks so much like the missing piece in the puzzle
 that it has been
questioned whether supersymmetry is really broken \cite{wit1}. But if it
isn't broken,
it is of course hard to explain why we do not see superpartners 
of the Standard Model fields \cite{wit3}.

The recently revived suggestion that we live on a 4--dimensional
brane that is embedded in a higher-dimensional bulk opens up
a new perspective and a way out (under an assumption stated in
section 2). It will be proposed below
 that supersymmetry is
indeed unbroken up to micrometer scales -- but only in the {\it bulk}
supergravity theory. By a simple argument,
supersymmetry breaking in the bulk at micrometer scale
is derived from supersymmetry breaking at the $TeV$ scale on
the brane, which carries the Standard Model fields.

Due to a mechanism first proposed in \cite{rub} and
re--invented in \cite{vv,schm}, the vacuum energy of the brane
fields is shown not to contribute to the $4d$ cosmological constant.
Rather, it is absorbed in the curvature transverse to the brane.
Only the vacuum energy of the bulk supergravity fields
is argued to contribute to the
$4d$ cosmological constant. 

This vacuum energy is estimated. The result is
a relation between the 
Planck mass $m_{Planck}$, the scale $m_{Brane-Susy}$ of supersymmetry
breaking on the brane,
 the scale $m_{Bulk-Susy}$
of supersymmetry breaking in the bulk sector
 and the Hubble expansion rate $H_0$: roughly,
\ba\log{m_{Planck}\over m_{Brane-Susy}}\ 
\sim\ {1\over2}\ \cdot\ \log{m_{Planck}\over m_{Bulk-Susy}}\ 
\sim\ {1\over4}\ \cdot\ \log({m_{Planck}\over H_0})\ \la{erika}\ea
(a more detailed relation is given in the text).
Based on the known values of $m_{Planck}$ and $H_0$,
this relation predicts gravitinos or other superpartners of
the supergravity multiplet with masses of order
$10^{-3} eV$
(which is inside experimental bounds \cite{zwi}) and a supersymmetry breaking
scale on the brane of $2-6$ $TeV$.
Conversely, based on the assumption that supersymmetry is restored in the 
Standard Model at energies not too much above the weak scale,
the relation explains the observed small value
of the cosmological constant.

The setup is introduced in section 2. In section 3 it is argued that
the $4d$ cosmological constant is zero as long as
the bulk supergravity theory is treated classically.
 In section 4 it is shown that the
quantum mechanical ground state energy of the supergravity
sector produces a cosmological constant that is within a few orders
of magnitude of
its observed value. Precise matching yields the predictions for
supersymmetry breaking in the Standard Model and in the bulk sector,
as explained in section 5.
Section 6 contains conclusions.

{}\subsection*{2. The setup}\scs{2}

We consider a 3--brane soliton that is
embedded in a ($4+n$)--dimensional bulk spacetime (figure 1). We assume that
the $n$ extra dimensions are compactified on some manifold ${\cal M}$.
The Standard Model fields are assumed to live only on the brane, while
gravity lives in the bulk. Let Vol(${\cal M}$) be the volume of
the compactification manifold, and let Vol(${\cal B}$) be the volume of
the ball ${\cal B}$ inside
${\cal M}$ that intersects with the brane.
Since we are going to consider non-supersymmetric branes
inside supersymmetric bulks (as, e.g., in \cite{shi}), we will identify the size (i.e. the thickness)
of the brane with
$l_{Brane-Susy}\sim m_{Brane-Susy}^{-1}$, the scale of
supersymmetry breaking in the Standard Model.
So roughly, Vol(${\cal B}$)
$\sim \ (l_{Brane-Susy})^n$. 

Similarly as in \cite{dim},
because the Einstein action is integrated over Vol(${\cal M}$)
while the Standard Model action is integrated only over 
Vol(${\cal B}$), the 4--dimensional Planck length $l_{Planck}$ is related
to $l_{Brane-Susy}$ by 
\ba({l_{Planck}\over l_{Brane-Susy}})^2\ =\ ({m_{Brane-Susy}\over m_{Planck}})^2\ \sim\ 
 {\hbox{Vol(${\cal B}$)}\over\hbox{Vol(${\cal M}$)}}\ \la{anna}\ea
(assuming a $(4+n)$--dimensional Newton constant of order one).

 \begin{figure}[htb]
 \vspace{9pt}
\vskip13cm
 \epsffile[1 1 0 0]{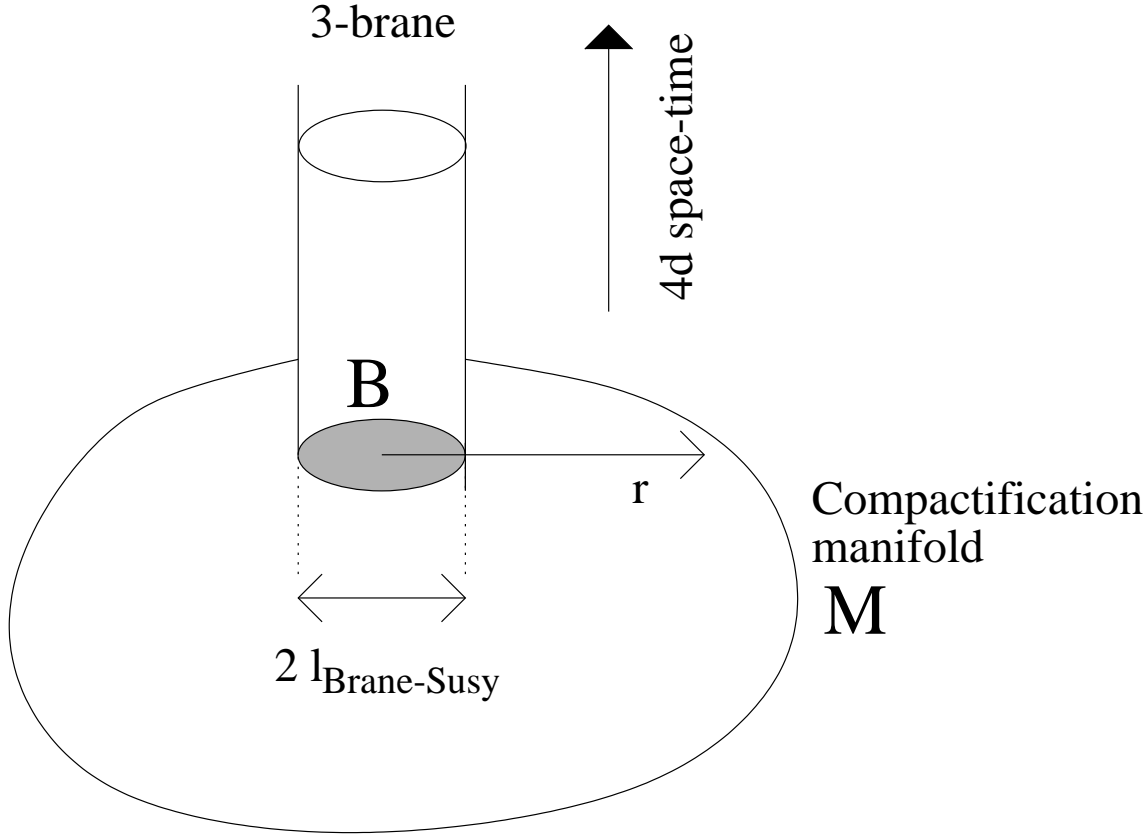}
\caption{A 3--brane in a $(4+n)$--dimensional embedding space.}
\end{figure}

Let us first consider the supersymmetric version of the story.
So we assume that we have a supersymmetric brane inside
a supersymmetric compactification manifold. In string theory, this is achieved
by considering a compactification on a Calabi-Yau 3--manifold 
that involves branes parallel to the 4--dimensional space--time,
as in \cite{ver1}.
At distances much larger than the size of the compactification
manifold, only a four--dimensional supersymmetric effective 
theory of Standard Model fields plus their superpartners coupled
to $4d$ supergravity is seen. 

For concreteness, we may assume a
 metric in the vicinity of the brane of the form
$$ds^2\ =\ dr^2\ +\ f(r)\ \hat g_{\mu\nu}dx^\mu dx^\nu\ +\ g(r)\ d\Omega_5^2\ $$
where $r$ denotes the distance from the brane, $x^\mu$
are the space-time coordinates parallel to the brane, $\hat g_{\mu\nu}$
is the $4d$ metric parallel to the brane, and $f(r),g(r)$
are some functions.
Supersymmetry of the effective $4d$ theory implies that
the $4d$ metric $\hat g_{\mu\nu}$
 is Ricci--flat (we are assuming that there are no
4--form gauge field strengths or expectation values of other
supergravity fields), 
i.e. the effective $4d$ cosmological
constant is zero. 

Now suppose that we cut out a region of radius $l_{Brane-Susy}$
around the brane. The basic assumption under which the arguments in the next section 
apply is that,  at the level of {\it classical} supergravity,
we can consistently  do the following: we can replace the supersymmetric brane soliton solution
by
a stable non--supersymmetric one (perhaps of the type of \cite{sen}),
 such that the bulk fields smoothly
connect to a solution at $r\ge l_{Brane-Susy}$
that does not break supersymmetry on the
$4d$ slices parallel to the brane. 

In other words, we assume that there are consistent string compactifications
that involve space-time filling stable   non-BPS branes,
such that $4$-dimensional supersymmetry is
unbroken away from the brane at least in the classical supergravity approximation.
  The construction of explicit examples
must be left for future work.

In the case of one extra dimension, examples 
 of supergravity solutions that smoothly interpolate
between a supersymmetric and a non--supersymmetric region
are the kink solutions of $5d$ gauged supergravity discussed in \cite{por,war}.

Supersymmetry is now broken not only in the bulk theory in the vicinity
of the brane. It is also broken in the
world--brane theory that contains the Standard Model fields
and lives in the non--supersymmetric gravitational background.
This will result in a brane vacuum energy of the order of
$(m_{Brane-Susy})^4$.

{}\subsection*{3. Classical Supergravity Approximation}\scs{3}

Let us first explain why the vacuum energy on the brane does not curve
the $4d$ metric $\hat g_{\mu\nu}$ parallel to the brane 
(i.e., why it does not create
a $4d$ cosmological constant) as long as the
bulk supergravity
theory is treated classically 
(see \cite{rub,vv,schm}). 
Although the bulk theory is treated
classically, the world--brane theory containing the 
Standard Model fields is assumed to be treated fully quantum mechanically.
Corrections from loops of the bulk fields are very interesting
and will be 
discussed in the next section.

The bulk has been separated into two regions: the non--supersymmetric
 neighborhood of the brane
$M^4\times{\cal B}$, where $M^4$ is the Minkowski space parallel to the brane;
and the supersymmetric region, i.e. the
 rest of the bulk $M^4\times({\cal M}-{\cal B})$.
The classical supergravity equations of motion can be solved separately for
each region, and can then be matched at their interface at $r=l_{Brane-Susy}$.

In the bulk  region, 
 the $4d$ metric $\hat g_{\mu\nu}$
 parallel to the brane
must still be Ricci--flat because of supersymmetry on the $4d$ slices parallel to the brane.

As for the brane region, there may be a singularity or horizon near the center.
Let us therefore restrict the discussion to the region
$\epsilon\le r\le l_{Brane-Susy}$, where $\epsilon$ is a cutoff that
hides the singularity or horizon. The issue of boundary conditions
at $r=\epsilon$ will be commented on below.

In this  non--supersymmetric brane  region,
 the brane is a source of vacuum energy $\rho$ of order $(m_{Brane-Susy})^4$
that arises from the world--brane fields. Let us
assume some distribution  $\rho(r)$  around $r=0$ 
 with
width of the order $l_{Brane-Susy}$. $\rho(r)$ enters the Einstein equations
like an $r$--dependent cosmological constant:
$$ R_{mn}-{1\over2}g_{mn}R= -\hat\lambda(r)\ g_{mn}\ $$
with
$$\hat\lambda(r)\ =\ 8\pi G\rho(r)\ -\ \lambda_{flux}(r)\ .$$
Here we have included another $r$--dependent contribution $\lambda_{flux}(r)$
that arises when the brane is a source of electric or magnetic flux.

For simplicity, we focus on the example of a single extra dimension,
assume a constant dilaton and 
neglect the other supergravity fields; the generalization is
straightforward. We make the metric ansatz
$$ds^2\ =\ dr^2\ +\ e^{2\alpha(r)}\hat g_{\mu\nu}dx^\mu dx^\nu.$$
In this ansatz, the $4d$ metric $\hat g$ is taken to be
$r$--independent. The 5--dimensional Ricci tensor can be written (cmp. with \cite{schm}): 
$$R^{(5)}_{\mu\nu}\ =\ \hat R^{(4)}_{\mu\nu}\ -\ \hat g_{\mu\nu}\ 
e^{2\alpha(r)}(\ddot\alpha+4\dot\alpha^2)\ $$
(a ``dot'' means ${d\over dr}$). Plugging this into the Einstein
equation for the 4-dimensional components ($\mu,\nu$) and using the
equation for the $(r,r)$ component to eliminate $\ddot\alpha$,
$$4(\ddot\alpha+\dot\alpha^2)=-{2\over3}\hat\lambda,$$
we obtain:
$$\hat R_{\mu\nu}^{(4)}\ =\ k^2\  \hat g_{\mu\nu}\ \ \ 
\ \ \ \hbox{where}\ \ \ k^2\ =\ e^{2\alpha}
({1\over2}\hat\lambda +  3\dot\alpha^2)$$
is an integration constant that is by definition the $4d$ cosmological 
constant ($k^2,\hat\lambda$ may be negative).
So the equations for $\alpha$ have a one--parameter family of solutions,
labelled by the constant $4d$ curvature $k^2$. However, matching
at $r=l_{Brane-Susy}$ to the solution in the supersymmetric region 
requires that we pick the solution that is Ricci--flat in $4d$, i.e. $k=0$.
For this solution, the vacuum energy on the brane is completely
absorbed by the warp factor 
$$\dot\alpha^2\ =\ -\ {1\over6}\ \hat\lambda(r)\ ,$$
and therefore does not curve the $4d$ metric parallel to the brane.
So the vacuum energy does not lead to a $4d$ cosmological constant.
For $n$ extra dimensions, the discussion is similar.

This is the mechanism of Rubakov and Shaposhnikov \cite{rub},
recently rediscovered in \cite{vv,schm}. We have supplemented it by a matching
condition at $r=l_{Brane-Susy}$ that picks out the solution with
vanishing $4d$ cosmological constant without fine--tuning.
This is a generalization of   the suggestion in \cite{ver2} of 
``supersymmetry on the Planck brane'' in the context of the Randall-Sundrum model.
Higher--order corrections 
will make the differential equations for $\alpha$ more complicated,
and there may be regions in parameter space where no solutions
exist \cite{kss}; let us assume that conditions are favorable
and solutions exist.

We have not discussed boundary conditions for $\alpha$ at the cutoff $r=\epsilon$,
where the supergravity approximation presumably breaks down.
However, whatever boundary conditions must be imposed -- the assumption
that they can be satisfied is part of the assumption that we have already
made in the previous section: that there are consistent string
compactifications that involve stable non-BPS branes and leave 4-dimensional
 supersymmetry
unbroken away from the brane  at the classical level.
Again, it remains to construct explicit examples.

{}\subsection*{4. Supergravity at One Loop}\scs{4}

Let us now go beyond the classical supergravity
approximation. This is the main new step taken in
this paper and it will lead to our numerical results.

 As mentioned in the introduction,
the ground state energies
$${\hbar\omega\over2}\ \ \ \hbox{with}\ \ \ \omega^2\ =\ k^2+m^2$$
of modes of light fields with momentum $k$
should give a quantum mechanical contribution to the cosmological constant.
We have already demonstrated that the ground state energy of the
Standard Model fields does not contribute to the $4d$ cosmological constant,
so it only remains to compute the vacuum energy produced by the bulk
supergravity fields: 
the gravitino, the dilaton, antisymmetric tensor fields
etc.

As long as supersymmetry is unbroken in the bulk, these vacuum energy
contributions cancel.
Now, breaking
supersymmetry in the region of the bulk 
near the brane also breaks supersymmetry 
in the effective $4d$ theory, obtained by integrating over the
compactification manifold ${\cal M}$.
But 
because of the small overlap of the wave functions of the supergravity
fields with the brane,
the mass scale $m_{Bulk-Susy}$ of supersymmetry breaking
in the bulk sector of the $4d$ effective theory will be suppressed
with respect to the scale of supersymmetry breaking 
 on the brane
 by the same volume factor that we already found in (\ref{anna}),
\ba({m_{Bulk-Susy}\over m_{Brane-Susy}})^2\ 
 \ \sim\
 {\hbox{Vol(${\cal B}$)}\over\hbox{Vol(${\cal M}$)}}\ \sim\ 
({m_{Brane-Susy}\over m_{Planck}})^2\ .
\la{beate}\ea

One way of seeing this is to consider a scalar field $\Phi$
in the supergravity
multiplet and  assume that it 
has a large mass of order
$m_{Brane-Susy}$ inside the region where supersymmetry is broken: 
we take its
$(4+n)$--dimensional Lagrangean to be of the form 
$$\partial_m\Phi\partial^m\Phi
 \ +\ \theta(r_{Brane-Susy}-r)\ m_{Brane-Susy}^2 \Phi^2 .$$
$\theta$ is the step function: $\theta(x)=0$ for $x<0$ and 
 $\theta(x)=1$ for $x\ge0$.
Integrating this Lagrangean over the compactification manifold,
the kinetic term acquires a prefactor Vol(${\cal M}$)
while the mass term only acquires a prefactor  Vol(${\cal B}$).
After normalizing $\Phi$ to have a standard kinetic term, 
its mass is
$$m^2_\Phi\ \sim\ m^2_{Brane-Susy}\ 
 {\hbox{Vol(${\cal B}$)}\over\hbox{Vol(${\cal M}$)}} \ .$$
This implies relation (\ref{beate}).\footnote{The relation 
$m_\Phi\sim{m^2_{Brane-Susy}\over m_{Planck}}$ could also have been 
derived without reference to branes; in this case
 the suppression factor is simply due to
the smallness of Newton's constant.}

So the hierarchy between the scales of supersymmetry breaking in the bulk
supergravity sector
 and supersymmetry breaking in the Standard Model
that lives on the brane is the same as the hierarchy
between the scale of supersymmetry breaking on the brane and the Planck scale.

{}Already in the introduction we have discussed the relation (\ref{diane})
 between
the momentum cutoff $\Lambda$ in the sum over vacuum energies and the
value of the cosmological constant $\lambda$. In the case of
$N$ massless bosonic 
propagating degrees of freedom, the relation changes to
\ba\lambda\ \sim\ N\ {\Lambda^4\over 2\pi}\ l_P^2\ .\la{ursula}\ea
$\lambda$ is related to the
Hubble expansion rate $H_0$ of the universe by
$$\lambda\ =\ 3\Omega_\Lambda\ H_0^2\ \ \ \hbox{with}\ 
\ \ \Omega_\Lambda\sim{2\over3}  $$
being the value suggested by observation \cite{bah}.
In a first estimate 
we may identify the cutoff 
$\Lambda$ in (\ref{ursula}) with the scale
$m_{Bulk-Susy}$ of supersymmetry breaking 
in the bulk.\footnote{The previous version of this paper
used this first estimate (i.e. it set $|Q|\sim1$ below) to 
suggest supersymmetry breaking scales of
$4-10$ $TeV$ on the brane and $10^{-3}-10^{-2}$ $eV$ in the
bulk (compare with section~5).}
 Then equation (\ref{ursula}) implies
(converting $l_{Planck}\sim m_{Planck}^{-1}$):
\ba
({m_{Bulk-Susy}\over m_{Planck}})^2\ \sim
 {6\Omega_\Lambda\pi\over N}({H_0\over m_{Bulk-Susy}})^2\ .\la{carola}\ea
A more precise calculation involves the various masses of order  $m_{Bulk-Susy}$
of the supergravity fields.
 Then $k$ in (\ref{diane})
is integrated not only up to $m_{Bulk-Susy}$, but up to
 $k_{max}\sim m_{Brane-Susy}$, which is the fundamental scale in our setup. 
Formula (\ref{diane}) generalizes to (see e.g. \cite{fer} for a discussion):
\ba
\lambda\ \sim\ {l_P^2\over2\pi}\ \times\ \sum_i(-1)^{F_i}\ \{\ k_{max}^4
\ -\ {1\over2}m_i^4\ \ln{k_{max}\over m_i}\ +\ ...\ \}\ ,
\ea
where $i$ counts the propagating degrees of freedom in
the supergravity multiplet,
$m_i$ are their masses of order $m_{Bulk-Susy}$
after supersymmetry breaking, and
$(-1)^{F_i}$ is $+1$ for bosonic and $-1$ for fermionic degrees of freedom.
The $k_{max}^4$ terms cancel since there is an equal number of bosons
and fermions.\footnote{A possible term of the form $k_{max}^2\sum_i
(-1)^{F_i}m_i^2$ that may appear in a more general calculation should
also vanish, since supersymmetry is broken {\it spontaneously}
by the non-supersymmetric soliton inside the supersymmetric bulk.}
The conclusion is then that $N$ in (\ref{carola}) is replaced by
\ba
N\ = Q\ \sum_i(-1)^{F_i}({m_i\over m_{Bulk-Susy}})^4\ \la{mia}\ea
with
$$ 
 \ Q\ \equiv\ -{1\over2}\ln{m_{Brane-Susy}\over m_{Bulk-Susy}}\ .$$
Together, with (\ref{beate}), (\ref{carola})  yields
the relation claimed in the introduction 
(where we have set $\Omega_\Lambda={2\over3}$ and roughly approximated
$({|N|\over4\pi})^{1\over2}$ by 1):
\ba\log({m_{Planck}\over m_{Brane-Susy}} )\ 
=\ {1\over2}\log({m_{Planck}\over m_{Bulk-Susy}})\ 
=\ {1\over4}\log({\sqrt{|N|\over6\Omega_\Lambda\pi}}{m_{Planck}\over H_0})\ 
\  \equiv \  L\ .\la{gloria}\ea

\subsection*{5. The Numbers}\scs{5}

Let us now plug in the numbers. We use
\ba
 m_{Planck}&\sim& 10^{19}\ GeV\\
{\sqrt{\lambda}}\ =\ {\sqrt {3\Omega_\Lambda}}H_0 &\sim& 2\cdot 10^{-33}\ eV\ea
What is $N$? Type IIB supergravity multiplets, e.g.,
have 128 bosonic and 128 fermionic degrees of freedom. 
Without going into details,
it seems safe to assume that $|{N\over Q}|$ in (\ref{mia})
is somewhere between 1 and 128. With $|Q|\sim20$, this gives
$${\sqrt{|N|\over2\pi}}\ \sim \ 2\ \ \ \hbox{to}\ \ \ 20\ $$
Since there may be other hidden factors of $\pi, {1\over2},$ etc.
that were missed by our crude analysis, the actual errors
in the relation (\ref{gloria}) may even be somewhat (but not much)
larger. Being optimistic about them, we infer that
$$L\ \sim\ 15.4\ \pm\ 0.2\ $$
This yields the predictions
\ba
m_{Brane-Susy}&\sim& \ \ 2\ TeV\ \ -\ \ 6\ TeV\\ 
m_{Bulk-Susy}&\sim& \ \ {1\over2}\cdot10^{-3}\ eV\ \ -\ \ {1\over2}\cdot10^{-2}\ eV\ .\ea
The example with the intermediate value of $L=15.4$ is
plotted in figure 2.

 \begin{figure}[htb]
 \vspace{9pt}
\vskip10cm
 \epsffile[1 1 0 0]{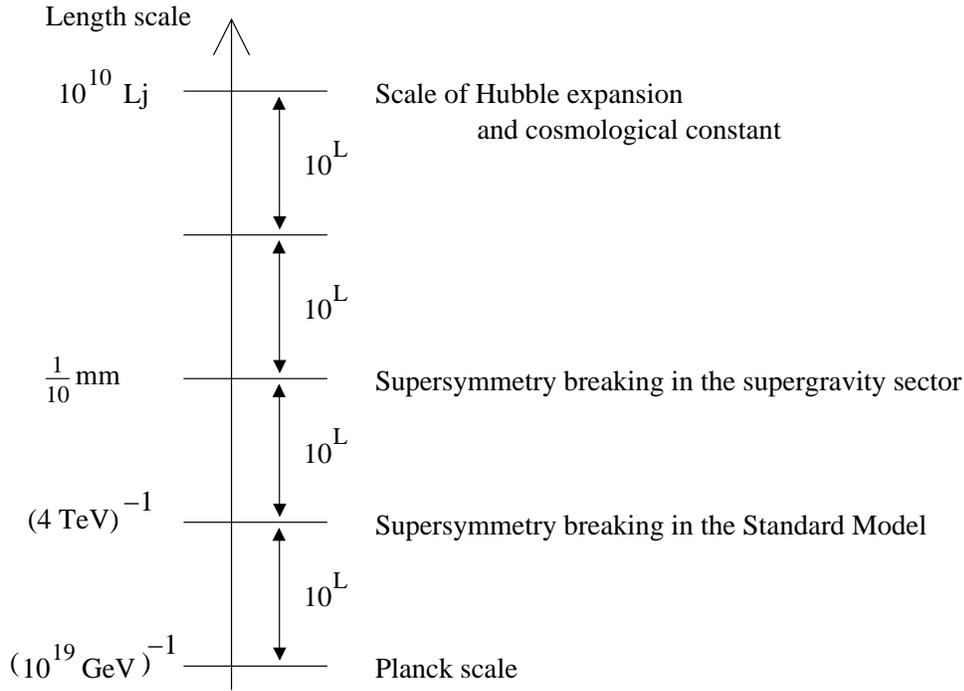}
\caption{The relation between hierarchies for the intermediate value $L=15.4$.}
\end{figure}

So we expect a mass of order $10^{-3}\ eV$ for the gravitino
or at least for some members of the supergravity
multiplet. This corresponds to 
a Compton wavelegth of the order of a fraction of a millimeter.
Similarly as in the case of millimeter--size 
extra dimensions \cite{dim},
the presence in the bulk of gravitinos or dilatons in the micrometer
range is  not ruled out by experiment: while
the brane physics (the Standard Model)
 has been probed down to the weak scale, the bulk physics (gravity) 
has only been probed down to centimeter scales.
The lower experimental bound on the gravitino mass appears to be 
only $10^{-5}eV$ \cite{zwi}.
The effects of these new fields 
might show up in short--distance measurements
of gravity in the $\mu m$ range in the near future \cite{exp}. 
 
Remarkably, the predicted
 scale of supersymmetry breaking in the Standard Model
is roughly where it is expected to be,
in order to insure that the running
coupling constants meet in supersymmetric
Grand Unification. This is very nontrivial; a priori
it could have come out many orders of magnitude off the mark.
Reversing the logic,
if we assume a probable scale of supersymmetry breaking
between 1 and 100 TeV, then we can predict the value of the cosmological 
 constant within a few orders of magnitude of the value that seems to
have been measured!

Let us finally note that our derivation and results apply just as well to
the case of a single extra dimension as in the
Horava--Witten model \cite{hor} or in the Randall--Sundrum model \cite{rs}.

\subsection*{Conclusion}

It seems that the proposal that we live on a 
 non--supersymmetric brane that is embedded in a
supersymmetric higher--dimensional string compactification
can explain the observed small value
of the cosmological constant, provided that the scale of 
 supersymmetry breaking  in the Standard Model is roughly  2--6 TeV. 
It remains to construct explicit examples of such compactifications and to
show that they are consistent. 

Our  explanation for the small cosmological constant
can be tested by searching for signs of
 a gravitino, a dilaton or other supergravity
fields with masses of order $10^{-3}$ $eV$.
We can thus look forward to a number of surprises
in future tests of Einstein gravity in the micrometer range.

\vskip5mm\noindent
{\bf Acknowledgements:}

\noindent
I thank W. Lerche and P. Mayr for discussions. I also thank
I. Antoniadis, R. Barbieri, S. Dimopoulos,
 S. Kachru, G. Veneziano, H. Verlinde and E. Witten
for comments on the previous version of this paper.
This work is supported in part by a Heisenberg fellowship of the DFG.

\vskip2mm
\noindent
{\bf Note:} After this paper appeared, I was notified by the authors of \cite{gab} 
that they mention on page 5 the possibility of using infinite  volume extra dimensions to
suppress the breaking of $5d$ bulk supersymmetry and the $5d$
 bulk  cosmological constant.
(Note however that we discuss $4d$ supersymmetry and the $4d$ cosmological constant on slices parallel to the brane.) 

\noindent
I also became aware of \cite{ant}, where non-supersymmetric branes
inside a supersymmetric bulk are constructed.

\end{document}